\documentclass[twocolumn]{aastex631}

\newcommand{\Msun}{$M_\odot$}

\begin{document}

\title{Swift Observations of \textit{Gaia} BH3}

\correspondingauthor{Boris Sbarufatti}
\email{boris.sbarufatti@inaf.it}

\author[0000-0001-6620-8347]{Boris Sbarufatti}
\affiliation{INAF–Osservatorio Astronomico di Brera \\ 
Via E. Bianchi 46 \\ 
Merate (LC), I-23807, Italy}

\author[0000-0003-4137-6541]{Renato Falomo}
\affiliation{INAF - Osservatorio Astronomico di Padova\\
Vicolo dell'Osservatorio 5\\ 
I-35122, Padova, Italy}

\author[0000-0002-0653-6207]{Aldo Treves}
\affiliation{ Universit\`a dell'Insubria, Dipartimento di Scienza e Alta Tecnologia\\
Via Valleggio 11\\ 
I-22100 Como, Italy}

\begin{abstract}

We present UV and X-ray observations, obtained with the \textit{Swift} Observatory, of \textit{Gaia} BH3 a binary system containing a 33 \Msun~ black hole discovered through \textit{Gaia} astrometry. 
  The system is well detected in all UV and optical filters (1700-6500 \AA). We compare our results with the modeling of the non collapsed component using synthetic stellar libraries, a good agreement is found with our UV observations

\end{abstract}

\keywords{Astrometric binary stars (79) --- Stellar mass black holes (1611) --- Ultraviolet astronomy(1736) --- G giant stars (557)}

\section{Introduction}\label{sec:intro}

    \textit{Gaia} BH3 (Gaia DR3 4318465066420528000) is an astrometric binary system containing a 33 \Msun ~ dormant black hole, discovered through astrometry by the \textit{Gaia} mission. The binary system has a period of 11.6 yr, eccentricity $e=0.729$, a semi-major axis of 16 AU, and a distance $d= 590$ pc. The visible component is a metal poor giant of mass $M=0.76$ \Msun, $T_{eff}$=5200 K, $\log(g)=2.929$, and [Fe/H]=-2.56 \citep{gaia2024}.
    
    \textit{Gaia} BH3 is a remarkable source because the mass of the unseen black hole component is significantly larger than that of black holes in X-ray bright systems, the most massive of which is Cyg X-1, $M=\sim20$ \Msun~ \citep{millerjones2021}. 
    
    In this work we report on UV and X-ray observations of \textit{Gaia} BH3 obtained with the Neil Gehrels \textit{Swift} Observatory.

\section{Swift Observations}\label{sec:obs}

\begin{figure*}
  \centering
  \includegraphics[width=\hsize]{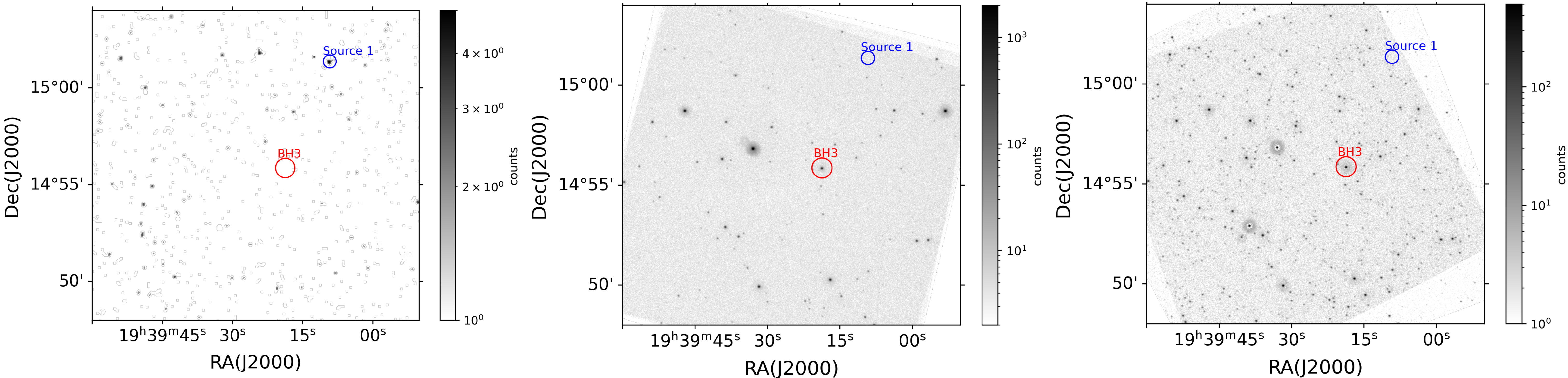}
     \caption{\textit{Swift} XRT (right), UVM2 (center) and U (left)of the field of BH3. XRT: the position of BH3 and the region used to estimate the upper limit are shown as a red circle. The blue circle shows the position of the only X-ray source serendipitously detected in the field.
     UVOT: the red circles show the position of BH3, the blue circle shows the position of the serendipitous X-ray source. The colorbars show the raw image counts.}
        \label{Fig:skyview}
\end{figure*}

    \textit{Swift} observed \textit{Gaia} BH3 in March-April 2024 within Target of Opportunity programs (ObsIDs 00016607001, PI J. Miller; and ObsID 00016607001, PI B. Sbarufatti), for a total exposure of $\sim 6$ ks. The observation journal and results are  reported in Tables \ref{Table:1}, \ref{Table:2}. X-ray Telescope \citep[XRT,][]{xrt1} and UV and Optical Telescope \citep[UVOT,][]{uvot} data were analyzed using  the standard software packages included with HEASOFT (version 6.33.2) and applying the relevant release of the calibration database. The XRT and UVOT U and UVM2 filter images are shown in figure \ref{Fig:skyview}.

\begin{table}
    \caption{Swift XRT observations of \textit{Gaia} BH3}             
    \label{Table:1}      
    \centering          
        \begin{tabular}{l c c  }    
            \hline\hline       
            Date & Exposure & 0.3-10 keV Flux \\
                 & \it (s)        & (\it $10^{-15}$ W~m$^{-2}$) \\
            \hline         
            2024-04-19  & 1379 & $<1.4$  \\
            2024-05-25  & 4555 & $<0.57$\\
            \hline                  
        \end{tabular}
\end{table}

    \subsection {X-ray Observations}\label{subsec:xrt}
    The XRT data of each observation were processed using  the \textit{xrtpipeline} task with standard calibration parameters. A search for X-ray sources consistent with the sky position of BH3 was performed using the XIMAGE detection algorithm on both data sets, with no source detected in either case, with a 90\% confidence level. We then combined the observations using XSELECT, and confirmed that no source coinciding with \textit{Gaia} BH3 is detected.
    Using a 30 pixel (70.8\arcsec) radius region for BH3 we estimate an upper limit of $8.1 \times 10^-3~ counts~ s^{-1} (4.0 \times  10^{-16}~W~m^{-2}$, assuming a power law spectrum with photon index $\Gamma=-2$). 
    This is consistent, but less constraining, with the results of  \textit{eRosita}  \citep[$F_X<1.2 \times 10^{-17}~W~m^{-2}$,][]{gilfanov2024}, and \textit{Chandra} \citep[$F_X<3.25 \times 10^{-18}~W~m^{-2}$,][]{cappelluti2024}.

   A X-ray source is serendipitously detected in the field, at position RA,DEC (J2000): 19h 39' 09.15", +15$^\circ$01' 21.7" with an error radius of 6.4\arcsec, at a rate of $3.3\pm0.8\times 10^{-3} ~counts ~s^{-1}$, equivalent to a flux of $1.6 \pm 0.4 \times 10^{-15}~W~m^{-2}$ assuming a power-law spectrum with  photon index $\Gamma=-2$). 
   No known source is reported by the SIMBAD database \citep{SIMBAD} or the NASA/IPAC Extragalactic Database (NED)\footnote{The NASA/IPAC Extragalactic Database (NED) is funded by the National Aeronautics and Space Administration and operated by the California Institute of Technology.} inside the XRT error circle.

    \subsection {UV observations}\label{subsec:uvot}
    In order to characterize the Optical/UV Spectral Energy Distribution (SED) of the source UVOT observations  were obtained, with a first pointing in April 2024, using all six standard filters  \citep{uvot2, uvot3}, and a second deeper pointing using only the UVM2 filter in May 2024 (see Table \ref{Table:2}).

    A list of all sources detected in the images was produced using the \textit{uvotdetect} tool to aid selection of a source and background region not contaminated by other sources. Multiple exposures in the same filter were co-added using the \textit{uvotimsum} task, and aperture photometry was obtained using \textit{uvotsource}, using a 5 $\arcsec$  radius aperture region centered at the \textit{Gaia} BH3 position, and an empty background region of 25 $\arcsec$ radius.  

    The source was detected in all UVOT six filters. Data of the B filter are strongly affected by saturation and therefore ignored. U and V filter data show some evidence of coincidence loss that was corrected using the appropriate corrections available in the UVOT calibration database while running \textit{uvotsource}, inspite of a higher uncertainty on the estimated fluxes for those bands.
    The UVW1 and UVW2 filters are contaminated by a red leak, that we corrected following the prescriptions suggested by \citet{Brown2010}. 
    The derived UV fluxes are given in Table \ref{Table:2}. The UVM2 data across the two epochs show no evidence of variability.

\begin{table*}
    \caption{Swift UVOT observations of \textit{Gaia} BH3}            
    \label{Table:2}      
    \centering          
        \begin{tabular}{l c c c c c} 
            Date & Exposure &Filter & Wavelength & Flux & Model Flux\\
             & \it (s) &  & \it  (nm)& \it  ($10^{-16}$ W~m$^{-2}$~nm$^{-1}$)& \it  ($10^{-16}$ W~m$^{-2}$~nm$^{-1}$)\\
            \hline\hline  
            2024-04-19 & 437 & UVW2   & 192.8 & $0.15\pm0.02$   & 0.11 \\
            2024-04-19 & 327 & UVM2   & 224.6 & $0.32\pm0.01$   & 0.30 \\
            2024-04-19 & 218 & UVW1   & 260.0 & $1.10\pm0.05$   & 1.02\\
            2024-04-19 & 109 & U      & 346.5 & $3.8\pm0.2$     & 3.59\\
            2024-04-19 & 109 & V      & 546.8 & $8.5\pm0.2$     & 8.21\\
            2024-05-25 & 4478& UVM2   & 224.6 & $0.310\pm0.007$ & 0.30\\
            \hline                  
        \end{tabular}
\end{table*}

\section{Synthetic Spectrum}\label{sec:SED}

\begin{figure}
   \centering
   \includegraphics[width=\hsize]{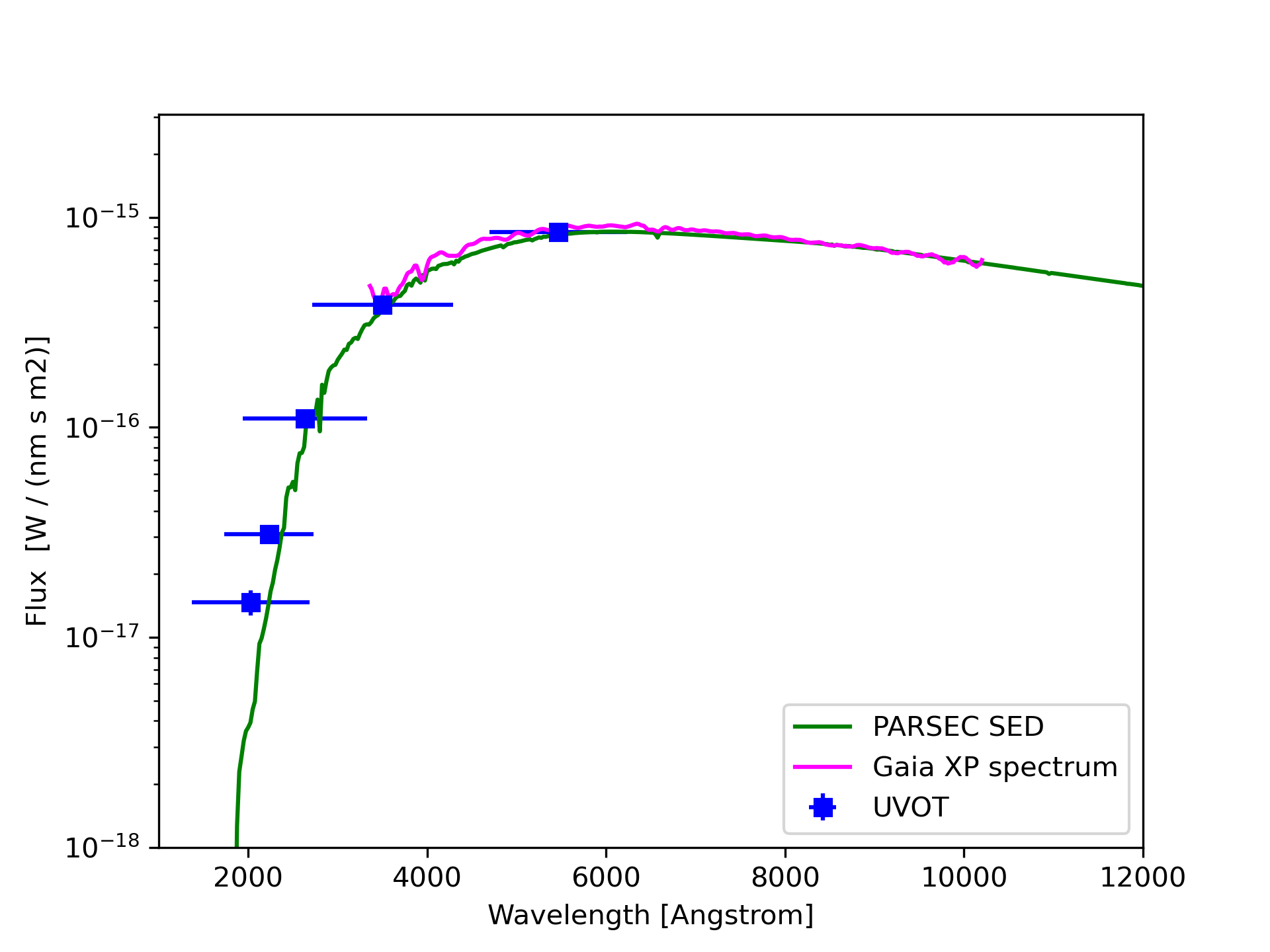}
      \caption{
    \textit{Gaia} BH3 Spectral Energy distribution in the UV/Optical bands. 
    green line: SED computed using the PARSEC libraries based on Gaia stellar parameters; Purple line: Gaia XP spectrum; blue squares: UVOT data (the horizontal bars in wavelength represent the FWHM of the UVOT filters, as reported by \citet{uvot2}.) 
              }
         \label{Fig:SED}
\end{figure}

In Fig.\ref{Fig:SED} we report, along with our UVOT data and the \textit{Gaia} XP spectrum \citep{gaiadr3}, a simulated spectrum, obtained using the YBC database \footnote{available at https://sec.center/YBC/index.html} and its interpolation code \citep{chen2019}, based on the PARSEC isochrones \citep{bressan2012}, assuming the physical parameters for the non collapsed component given by \citet{gaia2024} (see also Section \ref{sec:intro}), and using the PHOENIX spectral libraries \citep{allard95,allard97}. The reddening was taken into account following \citet{odonnell94, cardelli1989} with $A_V=0.76$. It is apparent that the good agreement between observations and synthetic spectrum reported by \citet{gaia2024} in the visible and near infrared bands extends also in the UV up to $\sim1600 ~ \text{\r{A}}$ ~ (Fig \ref{Fig:SED}). No indication of a UV excess is present. The predicted broad band fluxes are also reported in  Table \ref{Table:2}.

\section{Discussion}\label{sec:discussion}

\textit{Gaia} BH3 is the third black hole discovered through precise astrometric measurements by \textit{Gaia} 
\citep[for BH1 and BH2 see ][]{elbadry2023,elbadry2023b,tanikawa2023,chakrabarti2023}. The discoveries appear rather robust because the astrometric procedure is unaffected by the uncertainty on the inclination angle. 

These systems are much wider than those of classical X-ray binaries so that accretion and electromagnetic emission powered by Roche-lobe mass transfer can be excluded. A possible energy source could be the accretion of the interstellar medium. We quote the case of OGLE-2011-BLG-0462 \citep{sahu2022} where an isolated black was discovered in Hubble Space Telescope (HST) images through gravitational microlensing. Also in that case it was possible only to give an upper limit to the X-ray flux \citep{mereghetti2022}.  
Another likely accretion mechanism is an advection-dominated accretion flow due do wind-driven mass loss from the companion star, which has already been considered by \citet{cappelluti2024} for this system, but has also been proposed for BH1 and BH2 by \citet{rodriguez2024}

In the present work we focus on \textit{Gaia} BH3 and the UV band properties of the binary system in search of a possible UV contribution from the black hole to the total emission. Our observations do not show any significant UV flux in excess with the expected values from the model 
constrained at lower frequency. 
This is consistent with the expectations of \citet{cappelluti2024}, who proposed an advection-dominated accretion flow (ADAF) emission model for the accretion of \textit{Gaia} BH3.
In such a case the UV flux at $\sim2200~\text{\r{A}}$ ~ would be $<1$\% of the flux measured by us, which we attribute to the non collapsed component.

\bibliography{swift_bh3}{}
\bibliographystyle{aasjournal}

\begin{acknowledgements} 
    We wish to thank the referee for his constructive comments on this paper.
    We wish to thank Y. Chen for their valuable help in setting up the YBC/PARSEC software for the generation of stellar SED models. We thank the \textit{Swift} PI and the \textit{Swift} Science Operations Team for approving and supporting with the execution these Target of Opportunity observations.
    This work has made use of data from the European Space Agency (ESA) mission
{\it Gaia} (\url{https://www.cosmos.esa.int/gaia}), processed by the {\it Gaia}
Data Processing and Analysis Consortium (DPAC,
\url{https://www.cosmos.esa.int/web/gaia/dpac/consortium}). Funding for the DPAC
has been provided by national institutions, in particular the institutions
participating in the {\it Gaia} Multilateral Agreement.
This research has made use of the SIMBAD database,
operated at CDS, Strasbourg, France.
This research has made use of the NASA/IPAC Extragalactic Database (NED), which is funded by the National Aeronautics and Space Administration and operated by the California Institute of Technology.
\end{acknowledgements}

\end{document}